\documentclass[%
preprint,
superscriptaddress,
 amsmath,amssymb,
 aps,
pra,
]{revtex4-2}

\usepackage{graphicx}
\usepackage{dcolumn}
\usepackage{bm}
\usepackage{xcolor}

\begin{document}
\title{{\bf Is the end of Insight in sight?}}

\author{Jean-Michel Tucny}
\email{jeanmichel.tucny@uniroma3.it}
\affiliation{Department of Civil, Computer Science, and Aeronautical Technologies Engineering, Università degli Studi Roma Tre, via Vito Volterra 62, Rome, 00146, Italy}

\author{Mihir Durve}
\affiliation{Center for Life Nano- \& Neuro-Science, Italian Institute of Technology (IIT), viale Regina Elena 295, Rome, 00161, Italy}

\author{Sauro Succi}
\affiliation{Center for Life Nano- \& Neuro-Science, Italian Institute of Technology (IIT), viale Regina Elena 295, Rome, 00161, Italy}

\begin{abstract}
The rise of deep learning challenges the longstanding scientific ideal of insight—the human capacity to understand phenomena by uncovering underlying mechanisms. In many modern applications, accurate predictions no longer require interpretable models, prompting debate about whether explainability is a realistic or even meaningful goal. From our perspective in physics, we examine this tension through a concrete case study: a physics-informed neural network (PINN) trained on a rarefied gas dynamics problem governed by the Boltzmann equation. Despite the system’s clear structure and well-understood governing laws, the trained network’s weights resemble Gaussian-distributed random matrices, with no evident trace of the physical principles involved. This suggests that deep learning and traditional simulation may follow distinct cognitive paths to the same outcome—one grounded in mechanistic insight, the other in statistical interpolation. Our findings raise critical questions about the limits of explainable AI and whether interpretability can—or should—remain a universal standard in artificial reasoning.
\end{abstract}

\maketitle
\section{Introduction}

Recent advances in machine learning (ML), particularly through large language models (LLMs), have dramatically reshaped both science and society. These models now routinely tackle problems previously thought to be beyond reach, ranging from natural language understanding and protein folding to autonomous systems and symbolic reasoning \cite{vaswani2017attention, jumper2021highly, Nobel2024}. Such progress introduces a fundamentally different approach to scientific discovery — one based not on physical insight into underlying mechanisms, but on data-driven optimization through a dense web of parameters. While physics-informed constraints can improve convergence \cite{cai2021physics}, the learning process itself often remains opaque.

It no longer appears tenable to dismiss ML as a “glorified interpolator” or LLMs as “stochastic parrots” \cite{bender2021}. Instead, ML is beginning to challenge the very role of mechanistic understanding — or what has traditionally been called Insight — in scientific modeling. This tension raises the possibility of an "End of Insight" (EoI), a term coined by Strogatz \cite{EoI2007}, referring to the notion that certain scientific challenges may resist explanation in terms of simple governing principles, especially when they involve multiple interacting processes across vastly different scales in space and time.

This prospect is sad and perilous but plausible. Insight, as shaped by centuries of theory-driven physics, may not scale gracefully to problems such as epidemics, climate dynamics, or non-equilibrium statistical systems. ML, unconcerned with interpretability, may allow us to push the frontiers of knowledge in such domains — but without the perk of Insight and the intimate pleasure of "finding things out". This should not distract us from the fact that ML is still subject to a number of major limitations, especially when paired with the rising energy cost of training ever-larger models, a trajectory that raises concerns about sustainability and rapidly diminishing returns \cite{succi2024chatbots}.

In this paper, we explore both sides of this debate, using a concrete example: a deep neural network trained to solve a rarefied gas flow problem via physics-informed learning. While the network succeeds in replicating complex non-equilibrium behavior, we find little trace of physical structure in its trained weights — a result that challenges common expectations about interpretability and reinforces the divide between simulation and machine-learned representation.

Before turning to this example, we first recall the basic features of ML architectures, with an emphasis on their relation to complexity in physical systems.

\section{The basic ML procedure}

The basic idea of ML lies in approximating a $D$-dimensional output $y$ through recursive application of a nonlinear map \cite{lecun2015deep}. For a neural network (NN) with input $x$, $L$ hidden layers $z_1 \dots z_L$, each containing $N$ neurons, and an output layer $y$, the update chain reads:
\begin{eqnarray}
\label{FW}
z_0 =x\\
z_1 = f(W_1 x       -b_1),\; \dots  z_L = f(W_L z_{L-1} -b_L),\\
y   = z_{L+1} = f(W_{L+1} z_l -b_{L+1})
\end{eqnarray}
where $W_l$ are $N \times N$ weight matrices, $b_l$ are N-dimensional arrays of biases, and $f$ is a nonlinear activation function. At each layer, the output is often normalized $||z||=1$. The weights are updated via backpropagation, typically via a steepest descent: 
\begin{equation}
W' = W - \alpha \frac{\partial E}{\partial W}
\end{equation}
where $E[W] = ||y_T-y||$ is the loss function and $\alpha$ is the learning rate. 

\subsection{Taming complexity}

It is often claimed that, with enough data, ML can approximate virtually {\it any} target, whence the alleged demise of the scientific method \cite{anderson2008end,hasperue2015master}. Put down in such bombastic terms, the idea is readily debunked by general considerations on the physics of complex systems, see for instance \cite{coveney2016big,succi2019big}. Yet, ML does show remarkable proficiency in handling problems resistant to conventional modeling.

The natural question is: where does such magic come from?

To understand why, we briefly examine the three main boosters of Complexity: Nonlinearity, Nonlocality and Hyper-Dimensionality.

\subsubsection{Nonlinearity}

Nonlinear systems exhibit two distinguishing and far-reaching features: i) they do not respond proportionally to input, and ii) they transfer energy (information) across scales. This makes them erratic and hard to predict, but also capable of emergent phenomena—complex behavior arising from simple rules, biology being a goldmine of such instances. While physics has developed mathematical tools to handle nonlinearity, these are often overwhelmed when couplings become too strong across vast scales, with weather forecasting being a prominent example. ML can definitely help such methods stretch their limits. However, at present, there is no clear evidence that it can systematically outperform them, especially when precision is in high demand, as is usually the case for scientific applications \cite{coveney2024artificial}.

\subsubsection{Nonlocality}

In nonlocal systems, local behavior depends on distant states, often via long-range couplings. Although this interaction usually decays with the distance between the two regions, it cannot be ignored, no matter how far the interacting components are. A typical example from physics is classical gravitation, which is controlled by a potential decaying with the inverse power of the distance. The peculiarity of these systems is that they hardly reach a state of dynamic order known as "local equilibrium", usually controlled by a subset of "slow" variables living in a lower-dimensional manifold. Local equilibrium is the result of a neat scale separation between slow and fast variables, a feature which greatly simplifies the dynamics. Dynamics is notoriously much harder to capture than statistics and this is the reason why statistical physics is so effective in describing complex systems. With nonlocality in play, even statistical mechanics may remain hard to capture because of the aforementioned lack of scale separation between fast and slow modes. ML has shown promise in capturing such structures, particularly in identifying latent manifolds, though it remains an empirical rather than systematic approach \cite{strogatz2024nonlinear}.

\subsubsection{Hyper-dimensionality}

High-dimensional systems often exhibit the so-called curse of dimensionality, where the effective state space grows exponentially with each added variable. However, there is a subtler problem. Indeed, due to nonlinearities, heterogeneities and other factors, hyper-dimensional systems are usually very sparse: i.e. the "golden nuggets" are located in tiny corners of their state space. ML excels at exploiting this sparsity, adjusting billions of weights to traverse an immense function space and isolate meaningful correlations \cite{poggio2017and}.
 
The ML forward step described above sinks the problem into an ocean of weights, but the backwards step somewhat manages to fish it out by relentlessly chasing the golden nuggets, no matter how well hidden down into the ocean. To put it with Hinton, "Boltzmann machines are nice, but the real thing is the backpropagation algorithm".

For example, a DNN of width $N=10^3$ and depth $L=10^2$ contains $N^2 L=10^8$ weights and admits $P=10^{30}$ paths. Current leading-edge ML applications, such as DeepFold or LLMs, motoring the most powerful "ask-me-anything" chatbots, reach 100+ billions of weights, basically the number of neurons in our brain. Except that our brain operates on $\sim 20$ W, while these ML models require at least ten million times more.

These numbers unveil the magic behind ML: DNN duel the CoD face up, by unleashing an exponential number of paths and adjusting them to sensibly populate the sneaky regions where the golden nuggets are to be found. The rule is general, which is its blessing but also its curse. It is a blessing because it can escalate otherwise inaccessible peaks of complexity and, for the most ardent AI aficionados, hit the ultimate goal of "solving intelligence", thereby fixing all major problems in science and society (will it?) \cite{hassabis2017neuroscience}.
 
However, on a more prosaic and realistic ground, it is also a curse because it lacks Insight and quite possibly {\it just because of this}, it develops an insatiable appetite for energy. It is estimated that next-generation chatbots will approach the gigawatt power demand, more than most existing power plants. The comparison with the 20 watts of our brain is embarrassing. The question is whether the End of Insight also implies the end of the energetic resources of planet Earth, in which case one has probably to think twice before endorsing the "bigger is better" route undertaken by Big Tech companies.

The academic community is pursuing many efforts in this direction, if only with vastly inferior resources. In the following, we shall offer our modest token in this direction: the possibility of casting ML algorithms in terms of discrete dynamical systems. This would help sustainability, as with the identification of a DNN with a discrete dynamical system, weights immediately acquire a well-defined meaning in terms of generalized diffusion-advection processes, thereby opening the way to energy-leaner reduced representations \cite{NPDE}.
Let us describe the idea in more detail.

\section{Machine learning and discrete dynamical systems}

In a recent paper \cite{NPDE}, the ML procedure was formally reinterpreted as a discrete dynamical system in relaxation form: more precisely, as a time-discretized neural integro-equation (NIDE) of the form:

\begin{equation}
\label{REL}
\partial_t z = -\gamma (z-z^{eq})
\end{equation}
where $z=z(q,t)$ is the physical signal in spatial parameter $q$, and the local equilibrium
\begin{equation}
z^{eq} = f(Z)
\end{equation}
The mapping $Z$ is a shorthand for the shifted linear convolution
\begin{equation}
Z(q,t) = \int W(q,q') z(q') dq' - b(q)
\end{equation}
where $b(q)$ is the bias function.

The procedure is quite transparent, both conceptually and mathematically: the solution $z(q,t)$ is attracted to a local equilibrium $z^{eq}(q,t)$, the target of the procedure, which is the result of a nonlinear deformation, via the activation functional, of the convoluted signal $Z(q,t)$. The former is linear and non-local; hence, it implies scale mixing while leaving amplitudes untouched. The nonlinear deformation responds to a criterion of amplitude selection but leaves scales untouched. For instance, $tanh(Z)$ leaves small amplitudes unaffected and saturates the large ones on both sides. Rectified Linear Unit (ReLU), on the other hand, leaves positive signals unchanged and sets negative ones to zero. Hence, the signal is first non-locally linearly convoluted and then locally and nonlinearly deformed in amplitude. This sequence is key for ML schemes as universal interpolators, especially in high-dimensional spaces. Once again, the three boosters of complexity are fully accounted for.

A simple Euler time marching of the eq. (\ref{REL}), as combined with a suitable discretization of the "space" variable $q$ into a set of $N$ discrete nodes, delivers

\begin{equation}
\label{RELAX}
z_i(t+1) = (1-\omega) z_i(t) + \omega z_i^{eq}(t)
\end{equation}

where $\omega=\gamma  \Delta t$. Direct comparison with (\ref{REL}) shows that, with $\omega=1$, this is precisely the forward step of the ML procedure with $L=T/\Delta t$ layers and $N$ neurons per layer, with the initial condition $z(0)=x$ and output $y=z(T)$, $T$ being the time span of the evolution.  

Clearly, the result is highly dependent on the structure of the convolution kernel $W(q,q')$, whose discrete version is nothing but the weight matrix $W_{ij}$. In \cite{NPDE} it was noted that each kernel gives rise to a corresponding PDE and perhaps even low-order PDEs, such as advection-diffusion-reaction, with inhomogeneous, possibly time-dependent or even nonlinear coefficients, can give rise to pretty complex spatio-temporal patterns. Clearly, most common PDEs would lead to highly structured kernels, hence it was (naively) argued that inspection of real-life ML applications might show signatures of underlying structure. For instance, a simple advection-diffusion-equation in one spatial dimension would give rise to a tridiagonal-dominant weight matrix. The detection of such structural regularity in the weight matrices would offer a very valuable inroad to their explainability in the first place, let alone the energy savings resulting from a reduced set of weights.

The argument can be easily extended to more general PDEs, including strong inhomogeneities and nonlinearities, which could easily be implemented by convoluting local nonlinear combinations of the signal, that is:  

\begin{equation}
Z(q,t) = \int W(q,q') g(z(q')) dq'-b(q)
\end{equation}

where $g(z)$ is a local activation function, independent of $f$. 

For instance, by truncating the integral to the second moment, we would obtain $Z(q) = W_0(q) g(z) + W_1(q) \partial_x g(z) + W_2(q) \partial_{xx} g(z)$. In the above, the moments are defined as $W_k(q) = \int W(q,q') (q'-q)^k dq'$, and one may inspect their decay with increasing order to retain only a finite number in the sequence without seriously affecting the accuracy of the solution. The link between deep learning and PDEs is an active topic of research in the field \cite{han2018solving}.

\section{Inspecting the weights of a PINN application to rarefied gas dynamics}

The preceding considerations suggest that analyzing the weights of a trained network might offer insight into its internal logic, particularly when the problem is governed by a well-understood physical model. Let us test the idea by means of a concrete application. Recently, we trained a physics-informed neural network (PINN) on a body-force-driven rarefied gas flow through a 2D periodic array of cylinders in the laminar, isothermal and weakly compressible limit \cite{TUCNY2025102575}. This problem has a well-defined structure governed by the Boltzmann equation (BE).

A key parameter in rarefied gas dynamics is the Knudsen number $Kn = \lambda / D$, defined as the ratio of the molecular mean free path $\lambda$ to a characteristic length scale of the problem - in this case, the cylinder diameter $D$. The Knudsen number thus serves as a measure of rarefaction, characterizing the importance of non-equilibrium effects. In the continuum regime $(Kn \ll 1)$, the Navier-Stokes equations provide an accurate description of the flow. However, as $Kn$ increases, non-local effects due to the finite mean free path of molecules predominate, and momentum transfer is no longer only influenced by local velocity gradients. This nonlocal coupling is particularly evident in flows around curved surfaces.

\begin{figure}
\centering
\includegraphics[scale=0.3]{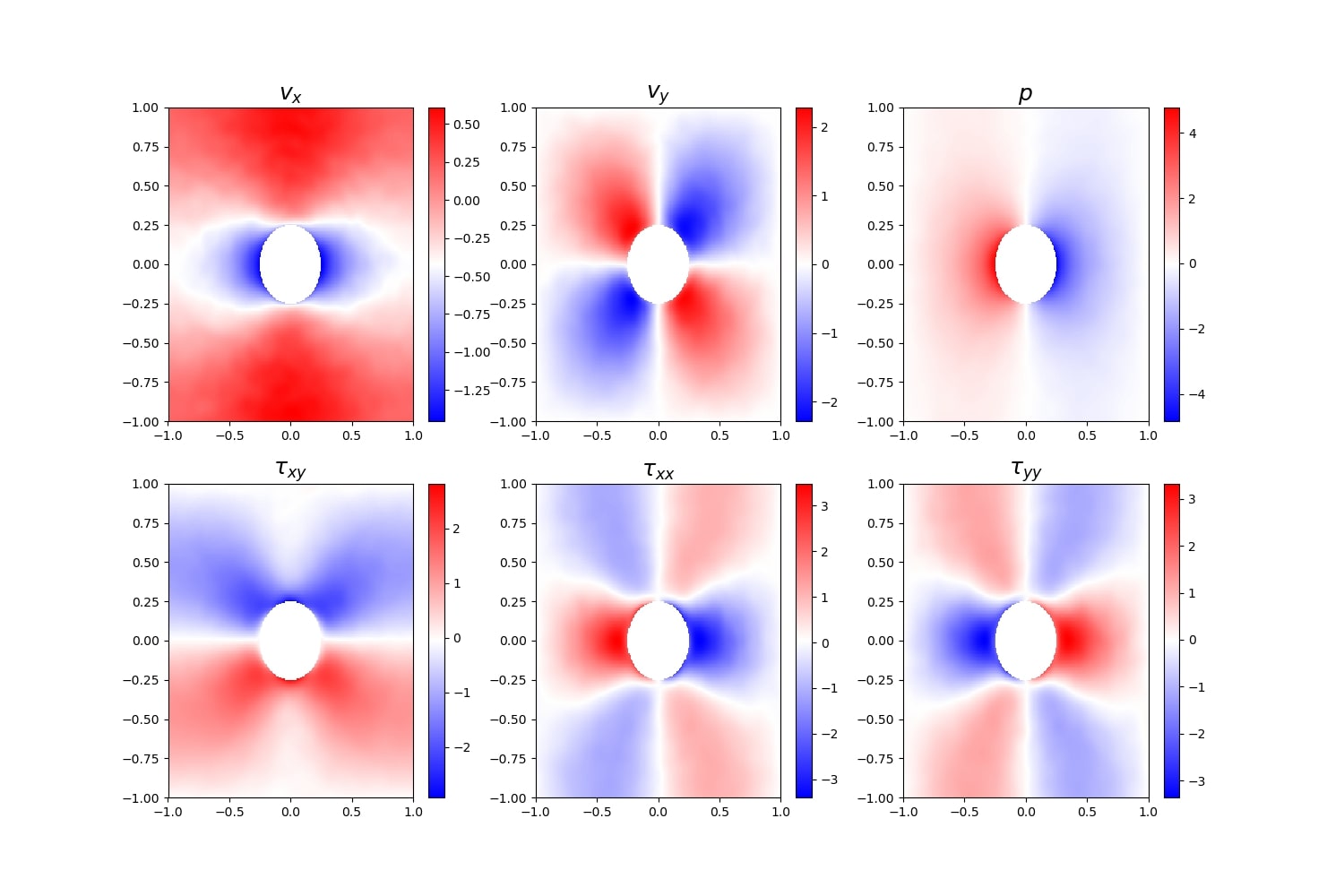}
\caption{Normalized macroscopic fields predicted by the PINN for $Kn = 1$. Each output variable was standardized across the full training domain in $x$, $y$ and $Kn$, resulting in zero mean and unit variance. This normalization, used to aid convergence, explains the presence of negative values in $v_x$ despite the imposed positive body force in the $x$-direction. See Ref. \cite{TUCNY2025102575} for details. \label{fig_MVPINN}}
\end{figure}

Motivated by these challenges, we designed a neural network that takes as input the spatial coordinates $(x_i,y_i), i=[1,256]^2$ and the Knudsen number $Kn$, and outputs the velocity components $u_x$, $u_y$, pressure $p$ and deviatoric components of the stress tensors $\tau_{xy}$, $\tau_{xx}$, $\tau_{yy}$, which are presented in Fig. \ref{fig_MVPINN}. This formulation captures the key physical quantities that characterize rarefied gas flows of industrial interest. While the input space is low-dimensional, the underlying physics is high-dimensional due to the dependence of transport properties on the full velocity distribution function. The network consists of a Fourier layer to impose periodic boundary conditions \cite{dong2021method,lu2021physics}, followed by nine hidden layers of 100 neurons with $tanh$ as an activation function. The neural network is regularized using the continuum equation and the Cauchy momentum equations.

The basic question is whether the trained network encodes recognizable physical structure. 

Before discussing the results, let us first show that our problem does exhibit the three key properties we described as where neural networks should excel. To this purpose, let us recall basic facts about the Boltzmann equation (BE).

\subsection{The Boltzmann equation}

This equation describes the dynamics of the probability density function $f(x,v;t)$, encoding the probability of finding a particle (atom, molecule) around position $x$ in space at time $t$ with molecular velocity $v$. In one dimension, and neglecting external forces: the BE reads as follows
\begin{equation}
\label{BE}
\partial_t f + v \partial_x f = Q(f,f)
\end{equation}
The left hand side represents the free streaming of the molecules, while the right hand encodes molecular collisions via a quadratic integral in velocity space involving the product $f(v)f(v')$ of two colliding particles with velocities $v$ and $v'$ at $(x,t)$. In full splendor:
\begin{equation}
Q(f,f) = \int P(v,w|v',w') [f(v)f(w)-f(v')f(w')] dw dv'dw'
\end{equation}
where $(v,w)$ and $(v',w')$ are the pre- and post-collisional velocities, and $P(v,w|v',w')$ is the probability of such collision; by micro-reversibility this is the same as the probability of the inverse collision from $(v',w')$ to $(v,w)$. In a way, $P$ can be interpreted as the weight kernel of the "kinetic Boltzmann machine", one which needs zero training since the physics supplies all the information it takes. In particular, the collision term is subject to mass-momentum-energy conservation laws, namely:
\begin{equation}
\int  Q(f,f) \lbrace 1,v,v^2 \rbrace dv = 0
\end{equation}

This structure embeds all three complexity boosters: nonlinearity via the quadratic collision term, nonlocality through the transport of information across space and velocity scales, and high dimensionality due to its formulation in six-dimensional phase space (plus time). While $Q$ is local in physical space, it is nonlocal in velocity space, and its competition with the streaming term drives the system toward or away from local equilibrium $f^{eq}$, depending on the Knudsen number $Kn$. In the hydrodynamic limit $(Kn \to 0)$, equilibrium dominates and the BE reduces to the Navier-Stokes equations. As $Kn$ increases, non-equilibrium effects emerge, and molecular-scale information propagates over macroscopic distances.

Even more relevant to macroscopic observables, integration of the BE over velocities yields transport equations that are simultaneously nonlinear and nonlocal in physical space, such as the familiar convective term $u \partial_x u$, with $u(x,t) = \int v f(x,v,t) dv / \int f dv$. This emergent structure underpins the complexity of fluid turbulence and forms the basis for the powerful lattice formulations of the BE \cite{succi2018lattice,PhysRevA.45.5771,rybicki2009prediction}.

\section{Learning the Boltzmann solutions via PINNs}

The PINN described earlier is trained by means of numerical data obtained by direct simulation Monte Carlo of the associated Boltzmann equation \cite{bird1976molecular}. Since the problem is exquisitely physical in nature, one may expect that the weights would somehow reflect the physical symmetries of the problem and the integro-differential nature of the Boltzmann equation. We then ask: does the trained network retain any visible trace of the problem’s physical structure?

\begin{figure}
\label{FIGPDF}
\includegraphics[scale=0.12]{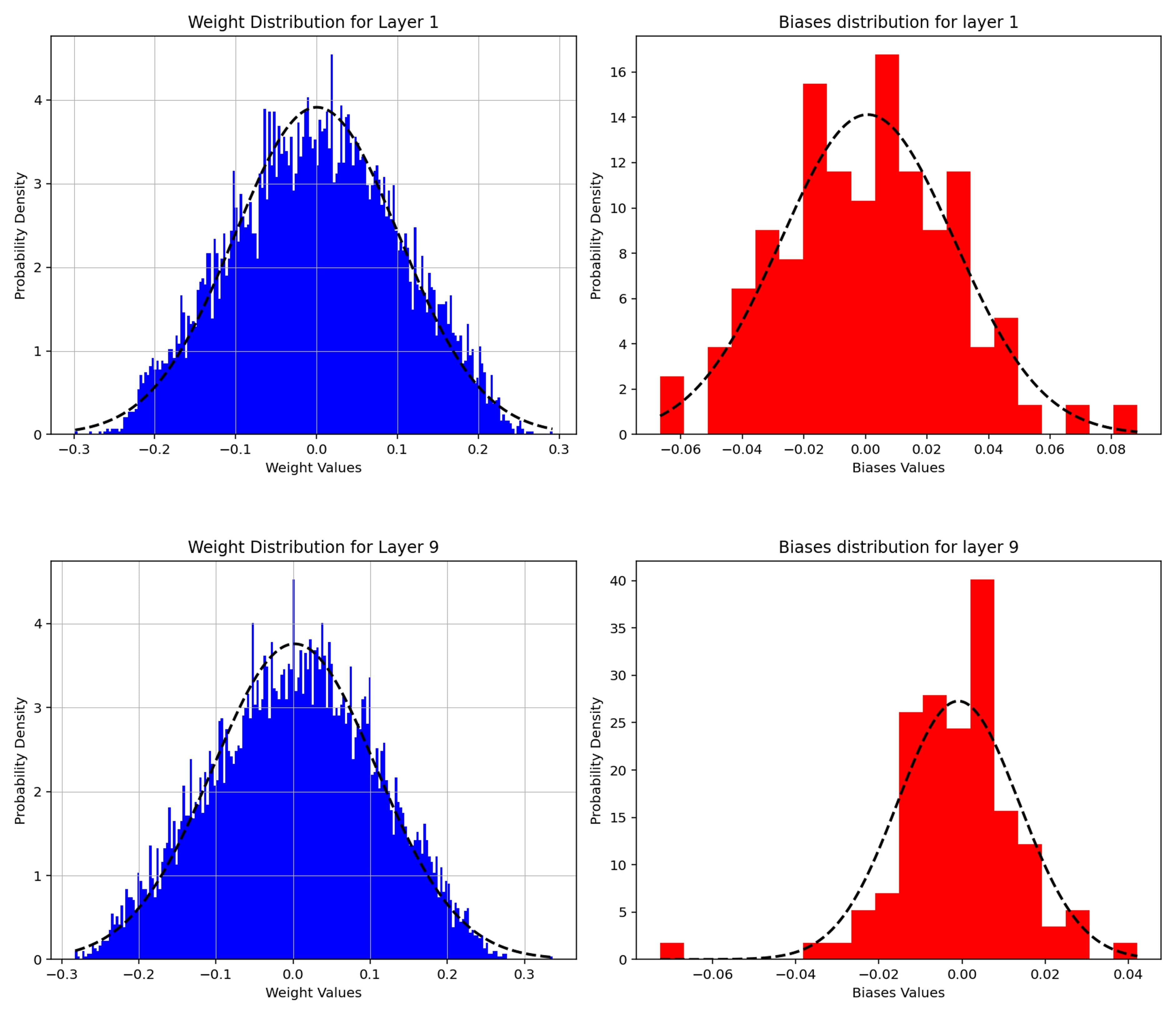}
\caption{PDFs of weights from the first and last deep layers. The symmetric shape indicates balance between excitatory and inhibitory contributions.\label{fig_weight_PD}}
\end{figure}

As seen in Fig.~\ref{fig_weight_PD}, the weight distributions in both shallow and deep layers closely resemble zero-mean Gaussians, suggesting no clear structural imprint from the underlying physics. While it could be argued that structure may emerge only in the composition of layers, the product of Gaussian-like matrices remains effectively random, making it unlikely that deeper inspection would reveal interpretable order.

One reason for this mismatch may be conceptual. The analogy between ML and discrete dynamical systems assumes an ordered metric structure for the discrete coordinates $q_i$, hence the weights $W_{ij}$. This is very restrictive in the context of ML at large, where the nodes host general input data and their interconnections do not generally bear any metric relevance. For our application, though, the lack of such metric structure is much less evident, since the input data is represented by the spatially structured pair of coordinates $x_i,y_i$. More precisely, on a regular spatial grid, the matrix associated with the discrete Boltzmann operator would exhibit a block-tridiagonal structure. The blocks might be random, reflecting the stochastic nature of the Monte Carlo method in velocity space, but they would still display a tridiagonal structure in configuration space.

Hence, it appears that the PINN-based deep learning procedure and the numerical solution of the Boltzmann equation offer two largely distinct paths to the dynamics of the rarefied gas system under inspection. 

\section{Tentative conclusions and outlook}

The inspection of the PINN trained on the rarefied gas problem suggests that its internal learning dynamics bear no direct (explainable) link to the physical structure of governing equations. Its weight matrices show little trace of the symmetries or integro-differential structure of the Boltzmann equation. In contrast, direct Monte Carlo simulations are firmly grounded in physical modeling. This points to two functionally equivalent, but epistemologically distinct, routes to knowledge.

Although based on a single yet representative example, our findings illustrate the possibility of knowledge gained without physical insight. That this occurs already for a problem of moderate complexity raises the question: beyond a certain threshold, might Insight become a luxury we can no longer afford?

This is not necessarily paradoxical. Explainability assumes compatibility with the scientific method, which is an artifact of the human brain. As modern ML systems increasingly diverge from biological architectures, there may be no a priori reason to expect that what they “solve” remains intelligible to us. Just as ML may be opaque to human understanding, the scientific method may be inaccessible to LLMs as we know them.

This might be disappointing at first glance, but it is not a problem per-se, as long as the lack of Insight does not translate into a reckless and potentially toxic use of ever larger and energy-devouring machines. 

Differently stated, lack of Insight does not mean that science has to abandon the pursuit of more economic and astute ways to advance our knowledge of complex systems, in the pursuit of an optimal combination of these two distinct paths to knowledge.

\section*{Author's disclosure on the use of AI for writing purposes}
The large language model ChatGPT-4o was used to assist in improving the clarity and conciseness of the English in this manuscript. The authors take full responsibility for the content and interpretations presented.

\section*{Acknowledgments}
SS is grateful to SISSA for financial support under the "Collaborations of Excellence" initiative and to the Simons Foundation for supporting several enriching visits. He also wishes to acknowledge many enlightening discussions with PV Coveney, A. Laio, D. Spergel and S. Strogatz. JMT is grateful to the FRQNT "Fonds de recherche du Québec - Nature et technologies (FRQNT)" for financial support (Research Scholarship No. 357484). SS and MD gratefully acknowledge funding by the European Union (EU) under the Horizon Europe research and innovation programme, EIC Pathfinder - grant No. 101187428 (iNSIGHT) and from the European Research Council ERC-PoC2 grant No. 101187935 (LBFAST).

\bibliography{sample}
\end{document}